# Unfolding of a diblock chain and its anomalous diffusion induced by active particles

Yi-qi Xia,[1] Zhuang-lin Shen,[1] Wen-de Tian,[1*] Kang Chen,[1*] and Yu-qiang Ma[2,1*]

1 Center for Soft Condensed Matter Physics & Interdisciplinary Research, Soochow University, Suzhou 215006, China
2 National Laboratory of Solid State Microstructures and Department of Physics, Nanjing University, Nanjing 210093, China
Corresponding authors: tianwende@suda.edu.cn (W. T); kangchen@suda.edu.cn (K.C.); myqiang@nju.edu.cn (Y. M.)

**Abstract**

We study the structural and dynamical behaviors of a diblock copolymer chain in a bath of active Brownian particles(ABPs) by extensive Brownian dynamics simulation in a two-dimensional model system. Specifically, the A block of chain is self-attractive, while the B block is self-repulsive. We find, beyond a threshold, the A block unfolds with a pattern like extracting a woolen string from a ball. The critical force decreases with the increase of the B block length ($N_B$) for short cases, then keeps a constant with further increase of $N_B$. In addition, we find a power law exists between the unfolding time, $t_{uf}$, of chain and active force, $F_a$, as well as $N_B$ with the relation $t_{uf} \propto F_a^{-1.1} N_B^{-1.34}$. Finally, we focus on the translational and rotational diffusion of chain, and find that both of them remain supper-diffusive at the long-time limit for small active forces due to an asymmetry distribution of ABPs. Our results open new routes for manipulating polymer's behaviors with ABPs.

**Introduction**

Understanding of single-chain behaviors of polymers is significantly important in polymer physics. In the case of single chain system, it is crucial to study and manipulate polymer conformations. On the dependence of environments, in the bulk or confined geometry, polymer chains can display the globule, coil, helix conformation.[1] In addition, the condensate structure of a DNA chain in multivalent cations shows a toroidal conformation.[2] The transition of conformations is also a fundamental problem being intensively studied in the past.[3] For example, the coil–globule transition is of importance in biology due to its presence in biological macromolecules such as proteins and DNA. The conformation transition of polymers can be experimentally manipulated by external stimuli such as pH[4], temperature[5], and mechanical force[6], which is of interest in biomedical engineering for controlled drug delivery.[7]

A polymer of single chain also exhibits amazing dynamical behaviors, especially, under the non-equilibrium conditions. For example, polymer in shear flow shows complex behaviors such as periodic elongation, relaxation, and tumbling.[8] Pronounced non-monotonic stretching of polymeric globules was also observed in the external flow field.[9] Recent experiments have fabricated variety of "supramolecular polymer"[10] or "colloidal polymer"[11], making it possible to study structural and dynamical behaviors of a single chain at a large length scale.

A passive polymer chain embedded in a fluid of active Brownian particles (ABPs), which can bring a non-equilibrium fluctuation to polymers, has been studied by various computer simulations, which found significant conformational changes of the chain because of the active environment.[12] ABPs are inherently non-equilibrium, consuming energy to produce motion.[13] They can generate the spatially varying mechanical pressure on chains.[14] For example, ABPs have been found to enhance the shape shift of circular chains[15] and trigger a modulational instability of semi-flexible polymers with pinning terminals.[14] Also, a temporarily stable hairpin conformation of semi-flexible polymer was found in the ABP bath and the polymer fluctuates between hairpin and stretch states in time.[16] These implied that the ABPs can be utilized to manipulate polymer conformations.

Our previous work[17] investigated the folding-unfolding transition of a collapsed chain in an active particle bath. Two unfolding mechanisms were proposed: shear tension and collision-induced melting. Here, we perform extensive Brownian dynamics simulations to investigate the structure and dynamics of an AB diblock chain immersed in an ABP bath in two dimensions. As we know, diblock copolymers have broad application in material modification and pharmaceutics.[18] The transition of their conformations is of interest in drug delivery. Particularly, in our system, the A block of the chain has a self-attractive property with a contracted conformation in solution without active particles, while B block favors the solvents and displays an expanded coil-like conformation. We study the conformation change of the chain induced by ABPs and pay attention to the influence of active force and length of B block. We also investigate the effect of active force on diffusive behaviors of chain. A novel unfolding pattern and anomalous diffusive character s of chain are observed.

**Model and Methods**

We employ the bead-spring model[19] to treat the AB diblock copolymer chain, immersed in an ABP bath in two dimensions, as shown in Fig.1. The A block composed of $N_A$ beads is self-attractive, and the B block with $N_B$ beads is self-repulsive, analogous to amphiphilic molecules in water solution[20]. The ABP is treated as a disk driven by a force $F_a$ along the direction $\hat{n}(t)$ which reorients under fluctuation.[21]

The interaction of self-attractive beads of A block is modelled by the smoothly-shifted Lennard-Jones(LJ)

potential, $U_{LJ}(r) = 4\varepsilon\left[\left(\frac{\sigma}{r}\right)^{12} - \left(\frac{\sigma}{r}\right)^{6}\right]$, which is cut off at $r_c = 2.5\sigma$. Here $\sigma$ is the diameter of each bead, $\varepsilon$ the interaction strength. Additionally, the LJ potential with $r_c = \sqrt[6]{2}\sigma$, i.e. the purely repulsive part, is adopted for all other non-bonded interactions including active particles and self-repulsive beads. The harmonic potential of successive beads on the chain is $U = k(r - r_0)^2$. $k$ is the spring constant, $r$ is the distance between the bonded beads and $r_0$ is the equilibrium bond length. Here we set $k = 2500 k_B T/\sigma^2$ and $r_0 = 0.98\sigma$.

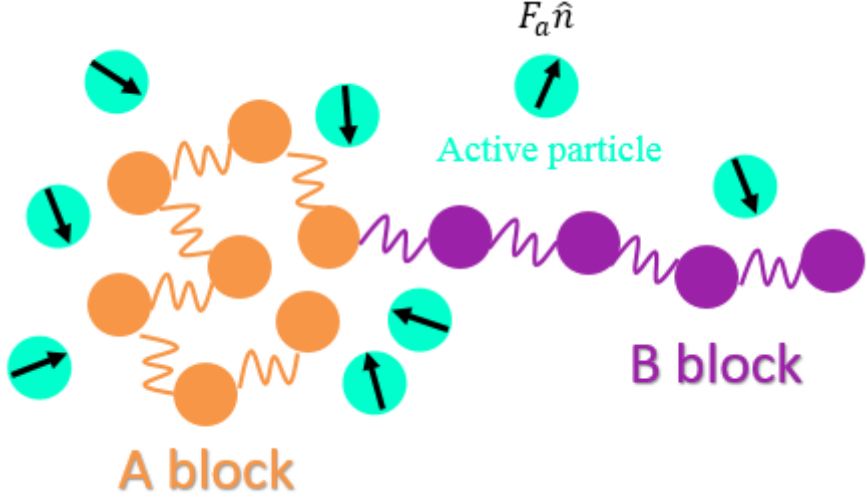

**Fig.1** Schematic illustration of model. The system is composed of an AB diblock chain and active Brownian particles. The monomers (orange disks) of A block are self-attractive, while the monomers (purple disks) of B block are self-repulsive. Both of them repel active particles (cyan disks).

The overdamped Langevin equation is used to describe the motion of ABPs and beads,

$$\dot{r}_i = (-\partial_r U + F_a \hat{\boldsymbol{n}}_i(\theta))/\gamma + \sqrt{2D_t}\boldsymbol{\eta}_i(t) \quad (1)$$

$$\dot{\theta}_i = \sqrt{2D_r}\xi_i(t) \quad (2)$$

where $r_i$ is the position of the $i$-th entity. Equation (1) governs the translational motion. For the chain beads, $F_a = 0$, $U_i$ is composed of both non-bonded potentials and bonded harmonic potentials, and only Eq. (1) is required. But for ABPs, $U_i$ contains only the non-bonded purely repulsive potentials and Eq. (2) is necessary to depict the coupled rotation of the driving direction. The translational friction coefficient $\gamma = k_B T/D_t$ where $D_t$ is the translational diffusion constant. $\boldsymbol{\eta}_i(t)$ is a Gaussian white noise induced by implicit solvent, which satisfies the fluctuation-dissipation theorem,[22] $\langle \eta_{j,\alpha}(t)\eta_{l,\beta}(t')\rangle = \delta_{jl}\delta_{\alpha\beta}\delta(t-t')$, where $\alpha$ and $\beta$ denote components of Cartesian coordinates. And the rotational noise $\xi_i$ is also Gaussian with zero mean and $\langle \xi_j(t)\xi_l(t')\rangle = \delta_{jl}\delta(t-t')$, where $D_r$ is the rotational diffusion constant.

We use the home-modified LAMMPS software[23] to perform the simulations with a periodic condition in both x and y directions. Square box with $L=200\sigma$ is adopted. Reduced units are used in the simulations by setting $\sigma=1$ and $k_BT=1$. The corresponding unit time, $\tau=\sigma^2/D_t$. We set $D_r = 1\times 10^{-3}$ as a constant, which is independent of $D_t$ similar to the previous work[15], so that the persistent time of ABPs is long enough to manifest their active features. Besides, we use $\varepsilon=10$ and set the friction coefficient $\gamma=10$. And, we control the motility of ABPs by varying the propelling force. All initial configurations were prepared at $F_a=0$ via a long time simulation. For each case, it was run by a minimum time of $1\times 10^4\tau$ with a time step, $\Delta t = 1\times 10^{-4}\tau$. The dimensionless active force is in unit of $k_BT/\sigma$.

**Results and Discussion**

Here we try to understand: 1) how does the chain unfold; 2) what are parameters that influence the chain unfolding; 3) what is the dynamics of chain in the ABP bath?

**1. Unfolding of chain**

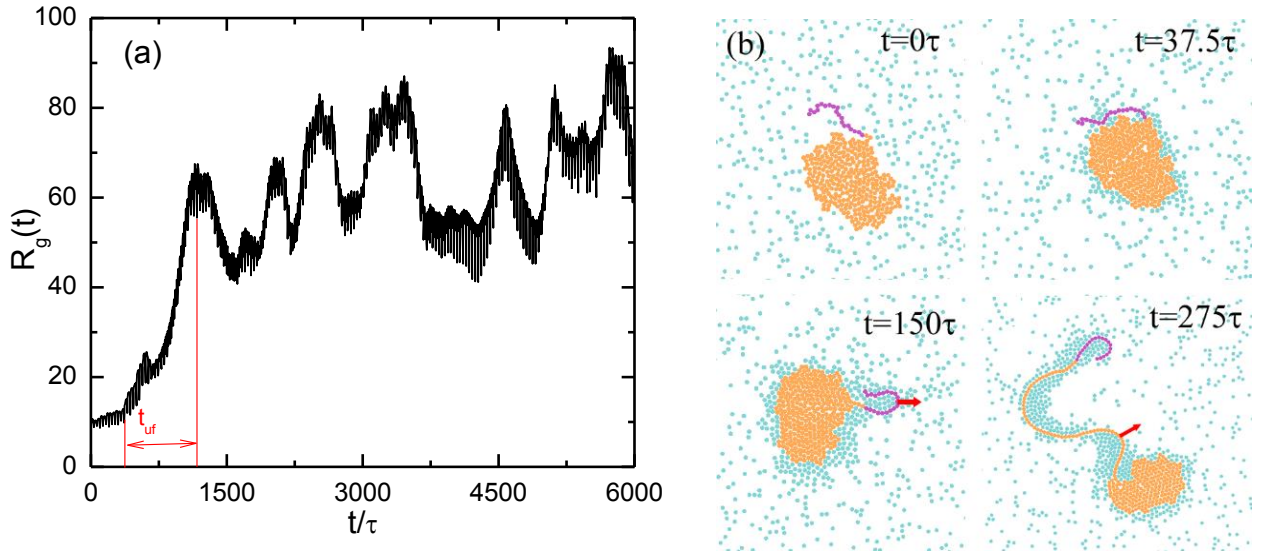

**Fig.2** (a) The time evolution of radius of gyration, $Rg = \frac{1}{N}\sqrt{\sum(r-r_{com})^2}$, of chain as a function of simulation time and (b) typical snapshots of chain configuration in the unfolding process of chain for $N_A$=300, $N_B$=20, $F_a$=7.5, and packing fraction of active particles, $\phi = 0.1$. $t_{uf}$ is the unfolding time of chain. The red arrow denotes the possible force direction induced by the active particle cluster.

We firstly investigate the behavior of the AB diblock chain of $N_A$=300 and $N_B$=20 immersed in the ABP bath with packing fraction, $\phi = 0.1$, and active force $F_a$=7.5. Radius of gyration of the chain was calculated and given in Fig.2a. It can be found that the size of chain increases at beginning, implying the unfolding of the chain. The unfolding velocity, at the outset, is low (Fig.2a) then it increases with the unfolding of chain. Finally the size keeps an average value of $R_g$=60, demonstrating the chain stays in an unfolding state.

Meanwhile, we can witness that the size of chain fluctuates largely, unlike a chain in good solution at equilibrium state.[24] Fig.2b shows the unfolding process of chain at $F_a$=7.5. At t=0τ, the A block of chain is in the collapsed state, and the B block of chain is in the coil state. After active force on ABPs was turned on, a number of active particles gradually aggregate around the chain (t=37.5τ) due to the self-trapping effect of active particles.[25] Then the B block of chain was stretched (t=150τ) due to an effective shear force imposed by active particles.[9] The shear force can induce a tension that stretches the chain. More beads of chain are extracted, more particles are adsorbed, which results in a stronger shear force that raises the unfolding speed. Subsequently, the A block of chain is extracted. Simultaneously, the ABPs accumulate around the unfolding part of chain (t=275τ) until the whole chain is opened up. The unfolding process is very interesting, in accompany with an arrangement of ABP distribution. Meanwhile the process is different with our previous work,[17] which found that the unfolding position is random. Here, we find the unfolding position is at the connecting point between A and B blocks, and the A block unfolds sequentially. To study parameters that influence the chain unfolding in detail, we define the unfolding time $t_{uf}$ equal to the time requirement from the initial to full unfolding of chain, as denoted in Fig.2a. It does not include the relaxation time before the chain starts to unfold and the time when the chain configuration is in the full stretch state.

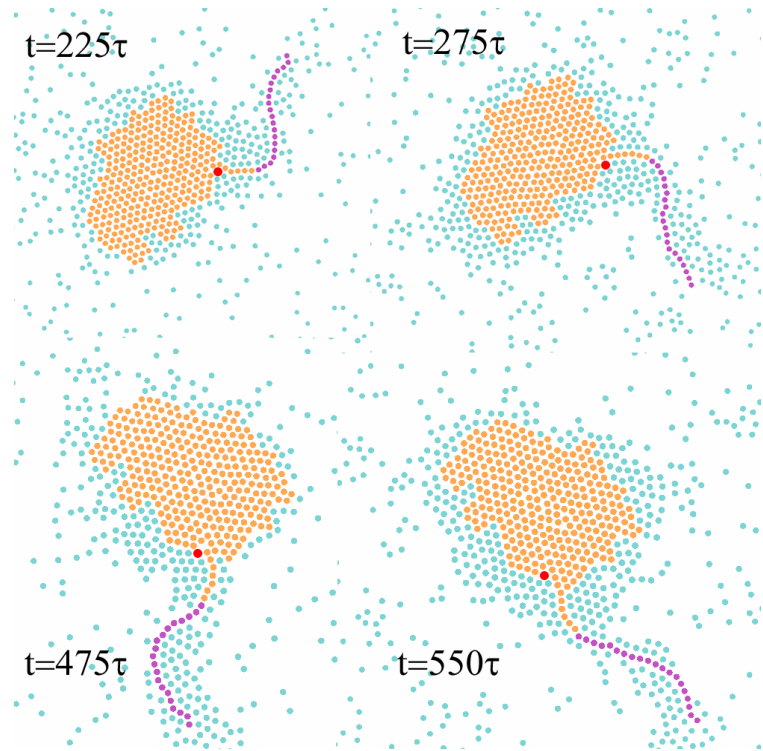

**Fig.3** Typical snapshots of chain unfolding through extrusive mechanism at $N_A$=300, $N_B$=20, $F_a$=6.25, and $\phi = 0.1$. The red point denotes the active particle that inserts into the collapsed region near the connecting point between A and B blocks.

Through a careful observation, we find another possible unfolding mechanism instead of the shear force at the beginning of unfolding, especially, for short $N_B$s. Near the critical force, it seems that the tensile force induced by ABPs could not extract the beads connecting the unfolding and folding part of chain. The beads of A block seem to be squeezed out by trapped particles through steric interactions, which is helpful for the particle layer formation around the chain. This is also why the unfolding speed is low at the beginning. As shown in Fig.3, after ABPs' accumulation, the B block are moving slowly from side to side, accompanied by the rotation of the cluster of A block. The swaying leads to a high-curvature region, which attracts many ABPs. Meanwhile, the swaying also causes the perturbation of relative distances or local structure of beads near the connecting point, which provides a possibility for the insertion of an ABP (red point in Fig.3). This extrusive mechanism hinders the chain collapse and promotes the chain unfolding.

## 2. Effect of active force

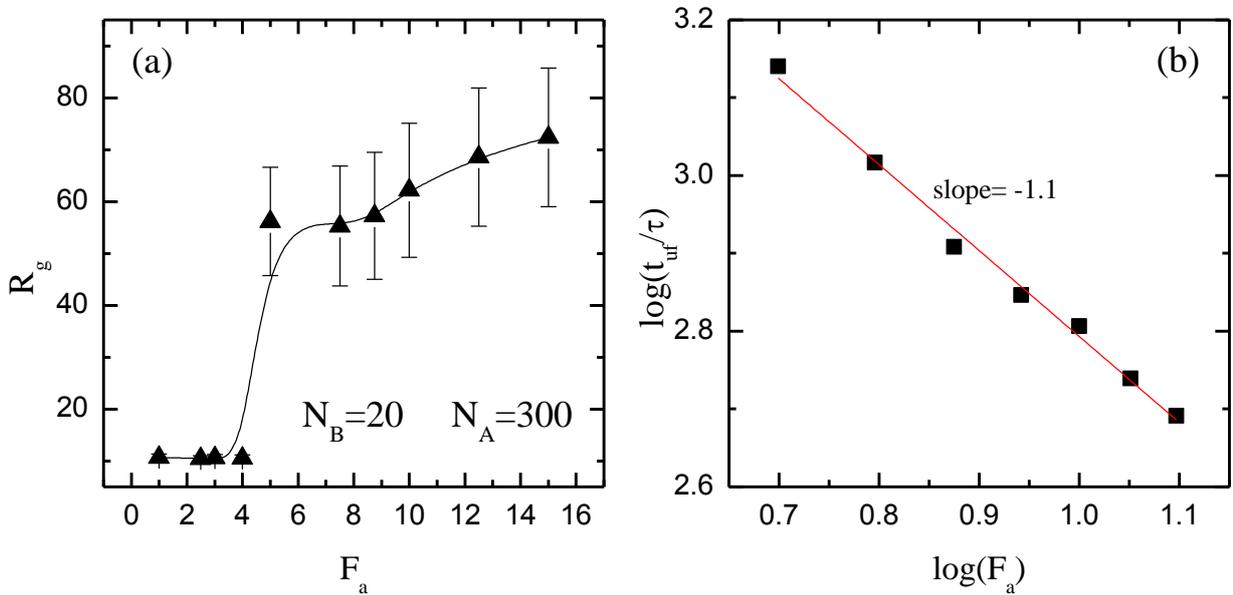

**Fig.4.** (a) $R_g$ as a function of active force, $F_a$ and (b) the unfolding time $t_{uf}$ vs $F_a$ at $N_A$ =300, $N_B$=20, and $\phi = 0.1$. The error bars were also shown.

We next turn to the influence of active force on the polymer conformation. It is evident from Fig.4a that there exhibits a critical force to unfold the chain. At low active force ($F_a$<=4.0), the chain (A block) remains in the collapsed (folding) state with $R_g \sim 10$. When the $F_a$>=5, the size of chain increases largely with $R_g >$ 50, indicating the chain stays in the unfolding state. With further increase in active force, the $R_g$ increases slowly with a large fluctuation. The folding-unfolding transition is very sharp with the increase of active force, like a first-order transition, similar to our previous work[17]. Obviously, a critical active force is needed to extract the A block of chain. Beyond the critical force, the unfolding time decreases with the increase of active force.

As shown in Fig.4b, we find a power-law relation between the unfolding time and active force, i.e., $t_{uf} \propto F_a^{-1.1}$. Because the length of A block is invariable, the above result reveals the average unfolding speed is proportional to $F_a^{1.1}$, not, $F_a$, implying there exists an enhanced effect on the tensile force as the active force increases. The flux of ABPs impinging on the chain is determined by $K_{on} \propto F_a$,[26] which signifies that the increase of $F_a$ causes not only the increase of effective tensile force resulting from each ABP but also the increase of number of impinging ABPs. Furthermore, the increase of active force makes it easy to unfold the chain at beginning. That's a possible reason that induces the average unfolding speed $\propto F_a^{1.1}$.

## 3. Effect of $N_B$

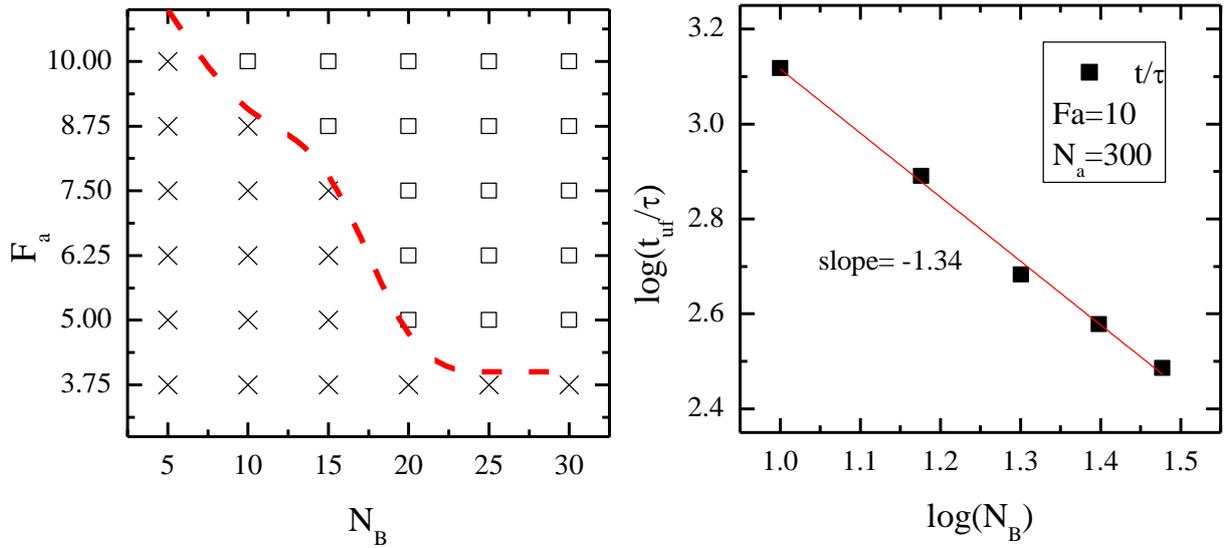

**Fig.5** (a) Phase diagram of chain configuration as functions of $N_B$ and $F_a$ at $N_A$=300. The red line is eye-guided boundary between folding state (×) and unfolding state (□) of A block of chain. (b) Plots of the unfolding time $t_{uf}$ vs $N_B$ for $F_a$=10 and $N_A$=300

Now, we pay attention to the influence of B-block length at $N_A$=300. It can be seen, from Fig.6a, the critical force decreases with the increase $N_B$ as it is shorter than 20. Then the critical force almost keeps a constant of 3.75 and does not change with further increase of $N_B$. Near the critical force, the unfolding is like a squeezing-out process with an extrusive mechanism. This mechanism depends on the accumulation of active particles and the local structure fluctuation resulting from $N_B$ block beating. The active force is a key factor that control the particle accumulation on the B block and around A block. Smaller active fore will cause the disappearance of particle aggregation and beating of $N_B$ block. That's why there exists a smallest critical force at large $N_B$s ($N_B$=20, 25, 30). On the other hand, the increase of $N_B$ infers a large tension that perturbs the local structure near A-B connecting point. This is why the critical force decreases with the increase of $N_B$ at small $N_B$s. Furthermore, we also take account of the relation between the unfolding time of chain and $N_B$. As plotted in Fig.6c, $t_{uf} \propto N_B^{-1.34}$ at $F_a$=10 and $N_A$=300.

## 4. Super-diffusion.

Finally, we study the effect of ABPs on the chain dynamics which is typically measured in terms of mean-square displacements (MSD). We compute the MSD, $\langle(\Delta r(t))^2\rangle = \langle(r(t) - r(0))^2\rangle$, for the center-of-mass translational diffusion of A block of the chain at various active forces for $N_A$=300, $N_B$=20. Meanwhile, we use the anomalous diffusion law, $\langle(\Delta r(t))^2\rangle \sim t^\beta$, to calculate the local scaling exponent $\beta^T(t) = d[\log(\langle(\Delta r(t))^2\rangle)]/d[\log(t)]$ along the ensemble averaged MSD trajectory. For Brownian motion $\beta^T(t) = 1$ at all times. It can be found from Fig.6a that at small active forces, where the chain is in the collapsed state, it shows a super-diffusive character at short and long times with $\beta^T(t) \sim 1.8$, as given in the inset of Fig.6a. At large active forces, the chain is in the unfolding state, it undergoes a super-diffusive behavior with $\beta^T(t) > 1$ before it finally reaches a normal diffusion regime at the long time limit. This is similar to the result of Kaiser and Löwen,[12] who studied a flexible self-avoiding polymer chain in an active particle bath.

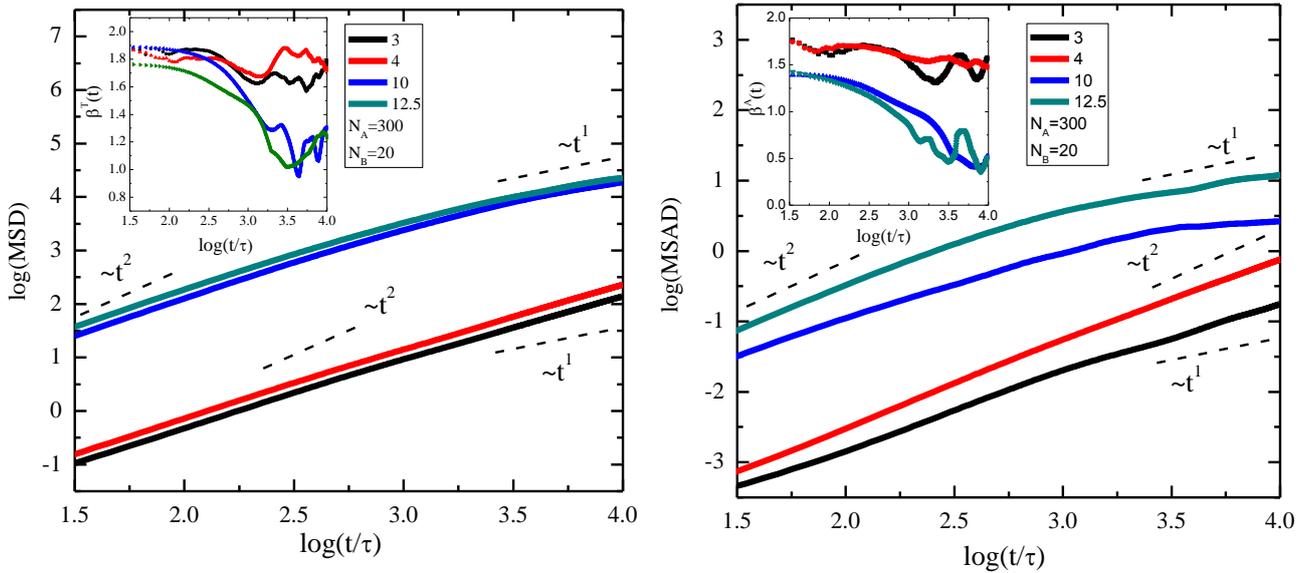

**Fig.6.** Log-log plots of (a) mean squared displacement (MSD) of center of mass of A block of the chain, and (b) mean squared angular displacement (MSAD) for various active forces at $N_A$=300, $N_B$=20. The dot lines (a guide for the eyes) have slope 2.0 and 1.0 respectively. The inset is scaling exponent $\beta(t)$.

We also focus on the mean-square angular displacement (MSAD), which is defined as $\langle(\Delta\theta(t))^2\rangle = \langle(\theta(t) - \theta(0))^2\rangle$, for its rotational behavior.[27] The $\theta(t)$ is an angle between x-axis and a vector from the center of mass of A block to its terminal. The local scaling exponent for rotation is defined as $\beta^A(t) = d[\log(\langle(\Delta\theta(t))^2\rangle)]/d[\log(t)]$. As given in Fig.6b, the MSAD shows super-diffusion for chain rotation at small active forces with $\beta^A(t) \sim 1.5$ at the long time limit. Interestingly, at large active forces, the exponent $\beta^A(t)$ decreases from ~1.4 to ~0.5, which reveals a transition of rotational motion from super-diffusion to sub-diffusion.

To explain why the translational and rotational motions of the chain remain super-diffusive at small active forces, we present typical snapshots of system with $N_A$=300, $N_B$=20, and $F_a$=4.0 in Fig.7. It can be found that there exists a stably asymmetry particle distribution around A block. More particles accumulate in the concave regions, especially, around the connecting region between A and B blocks due to the trapping mechanism. This asymmetry distribution of particles leads to a persistent pushing force on A block. Additionally, active particles favor aggregation in the high curve region of B block. This will result in a torque on the A cluster, which brings the rotation of A block to be super-diffusive.

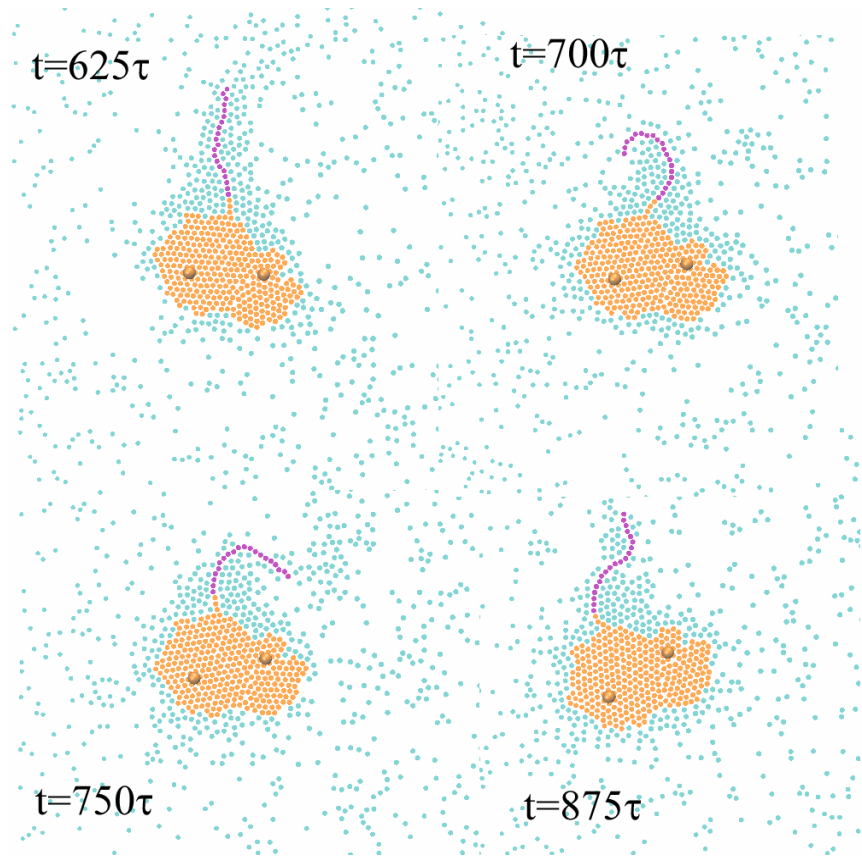

**Fig.7**: Time evolution of polymer trajectory for $N_A$=300, $N_B$=20, $F_a$=4.0. Two big spheres were given to display the rotation of the collapsed A block.

**Conclusion**

In summary, we employ Brownian dynamics to study the conformation and dynamics of an AB diblock chain embedded in the active particle bath. We find a critical force is needed to extract A block which is in a collapsed state when the active force on ABPs is turned off. Above the critical force, the chain unfolds with a pattern similar to extracting a woolen string from a ball. The monomers of A block are squeezed out at the beginning due to an extrusive mechanism and then the chain is opened up through the effective tensile force (due to impinging ABPs) from unfolding part. Then, we focus on the effects of length of B block and find that the critical force decreases with the increase of $N_B$, then it stays ~3.75 with further increase of $N_B$. In addition, there exist power laws between the unfolding time and active force with $t_{uf} \propto F_a^{-1.1}$, and $N_B$ with the relation $t_{uf} \propto N_B^{-1.34}$. Finally, we pay attention to the translation and rotation of chain. We find the chain shows a supper-diffusive behavior at the long time limit for small active forces, which can be explained by an asymmetry distribution of active particles. At large force, the rotation of chain displays a transition from super-diffusion at short times to sub-diffusion at long times. Our results open new routes for manipulating polymer and, possibly, for a new generation of polymer-based drug carriers.

Note that our model does not take account of the long-ranged hydrodynamics interactions,[28] which can tune collective behaviors of active agents, such as those in a 2D diffusion of micron-sized spheres driven by swimming bacteria.[29] Nevertheless, we believe that the main features of our findings stay valid, qualitatively, in real systems.

**Acknowledgment**

This work was supported by the National Science Foundation of China (NSFC). W. Tian acknowledge financial support from NSFC Nos. 21474074 and 21674078. K. Chen acknowledge financial support from NSFC Nos. 21574096, 21774091, and 21374073.

Reference:

1   G. R. Strobl, *The physics of polymers: concepts for understanding their structures and behavior*, Springer, Berlin ; New York, 3rd. rev. and expanded ed., 2007.
2   L. R. Brewer, M. Corzett and R. Balhorn, *Science*, 1999, **286**, 120–123.
3   V. B. Teif and K. Bohinc, *Prog. Biophys. Mol. Biol.*, 2011, **105**, 208–222.
4   E. Sherman and G. Haran, *Proc. Natl. Acad. Sci.*, 2006, **103**, 11539–11543.
5   A. I. N., Yiwei Fei, D. L. H. And and S. C. G., *Macromolecules*, 2007, **40**, 2559–2567.
6   J. Jaspe and S. J. Hagen, *Biophys. J.*, 2006, **91**, 3415–3424.
7   W. Tian and Y. Ma, *Chem Soc Rev*, 2013, **42**, 705–727.
8   A. Alexander-Katz and R. R. Netz, *Macromolecules*, 2008, **41**, 3363–3374.
9   A. Alexander-Katz, M. F. Schneider, S. W. Schneider, A. Wixforth and R. R. Netz, *Phys. Rev. Lett.*, 2006, **97**, 138101.
10  J.-M. Lehn, *Polym. Int.*, 2002, **51**, 825–839.
11  D. J. Ashton, R. L. Jack and N. B. Wilding, *Soft Matter*, 2013, **9**, 9661.
12  A. Kaiser and H. Löwen, *J. Chem. Phys.*, 2014, **141**, 044903.
13  C. Bechinger, R. Di. Leonardo, H. Löwen, C. Reichhardt, G.Volpe and G.Volpe, *Rev. Mod. Phys.*, 2016, **88**, 045006.
14  N.Nikola, A.P.Solon, Y. Kafri, M. Kardar, J. Tailleur and R. Voituriez, *Phys. Rev. Lett.*, 2016, **117**, 098001.
15  W.-D. Tian, Y. Gu, Y.-K. Guo and K. Chen, *Chin. Phys. B*, 2017, **26**, 100502.
16  J. Harder, C. Valeriani and A. Cacciuto, *Phys. Rev. E.*, 2015, 90, 062312.
17  Y. Xia, W. Tian, K. Chen and Y. Ma, *ArXiv180512292*.
18  M. A. C. Stuart, W. T. S. Huck, J. Genzer, M. Müller, C. Ober, M. Stamm, G. B. Sukhorukov, I. Szleifer, V. V. Tsukruk, M. Urban, F. Winnik, S. Zauscher, I. Luzinov and S. Minko, *Nat. Mater.*, 2010, **9**, 101–113.

19  P. E. Rouse, *J. Chem. Phys.*, 1953, **21**, 1272–1280.

20  R. Sinibaldi, C. Casieri, S. Melchionna, G. Onori, A. L. Segre, S. Viel, L. Mannina and F. De Luca, *J. Phys. Chem. B*, 2006, **110**, 8885–8892.

21  F. Schweitzer, in *Stochastic Processes in Physics, Chemistry, and Biology*, eds. J. A. Freund and T. Pöschel, Springer Berlin Heidelberg, Berlin, Heidelberg, 2000, vol. 557, pp. 97–106.

22  J. Weber, *Phys. Rev.*, 1956, **101**, 1620–1626.

23  S. Plimpton, *J. Comput. Phys.*, 1995, **117**, 1–19.

24  G. Cheng, W. W. Graessley and Y. B. Melnichenko, *Phys. Rev. Lett.*, 2009, **102**, 157801.

25  A. Kaiser, H. H. Wensink and H. Löwen, *Phys. Rev. Lett.*, 2012, **108**, 268307.

26  G. S. Redner, M. F. Hagan and A. Baskaran, *Phys. Rev. Lett.*, 2013, **110**, 055701.

27  J. M. Kim and C. Baig, *Sci. Rep.*, 2016, **6**, 19127.

28  M. C. Marchetti, J. F. Joanny, S. Ramaswamy, T. B. Liverpool, J. Prost, M. Rao and R. A. Simha, *Rev. Mod. Phys.*, 2013, **85**, 1143.

29  X.-L. Wu and A. Libchaber, *Phys. Rev. Lett.*, 2000, **84**, 3017.